\documentclass[apj]{emulateapj}

\usepackage{amsmath}
\usepackage{apjfonts}
\usepackage{amssymb}
\usepackage{natbib}
\usepackage{mathrsfs}
\usepackage{color}
\def\citea#1{\citep{#1}}
\def\citea#1{}

\begin{document}

\shorttitle{Detecting binary compact-object mergers with gravitational waves}

\title{Detecting binary compact-object mergers with gravitational waves:\\ Understanding and Improving the sensitivity of the PyCBC search}

\author{
Alexander H. Nitz\altaffilmark{1,2}, Thomas Dent\altaffilmark{1}, Tito Dal Canton\altaffilmark{1,3}, Stephen Fairhurst\altaffilmark{4}, and Duncan A. Brown\altaffilmark{2,5}}

\altaffiltext{1}{Max-Planck-Institut f{\"u}r Gravitationsphysik (Albert-Einstein-Institut), D-30167 Hannover, Germany}
\altaffiltext{2}{Department of Physics, Syracuse University, Syracuse NY 13244, USA}
\altaffiltext{3}{NASA Postdoctoral Program Fellow, Goddard Space Flight Center, Greenbelt, MD 20771, USA}
\altaffiltext{4}{School of Physics and Astronomy, Cardiff University, Cardiff, UK} 
\altaffiltext{5}{Kavli Institute for Theoretical Physics, University of California at Santa Barbara, Santa Barbara, CA 93106, USA}

\begin{abstract}
We present an improved search for binary compact-object mergers using a network of 
ground-based gravitational-wave detectors.  We model a 
volumetric, isotropic source population and incorporate the resulting 
distribution over signal amplitude, time delay, and coalescence phase into the 
ranking of candidate events. We describe an improved modeling of
the background distribution, and demonstrate incorporating a prior model of the 
binary mass distribution in the ranking of candidate events. 
We find a $\sim 10\%$ and $\sim 20\%$ increase in detection volume 
for simulated binary neutron star and neutron star--binary black hole systems, respectively, 
corresponding to a reduction of the false alarm rates assigned to signals by between 
one and two orders of magnitude.
\end{abstract}


\maketitle

\section{Introduction}
\label{s:intro}

The observation of binary black hole
mergers~\citep{Abbott:2016blz,Abbott:2016nmj,Abbott:2017vtc} by
the Advanced Laser Interferometer Gravitational-wave Observatory
(LIGO)~\citep{TheLIGOScientific:2016agk} has provided a new way of exploring 
the origin and evolution of compact objects. The measured masses, spins, and orientations of merging binaries 
can tell us about the nature and formation of compact-object remnants~\citep{TheLIGOScientific:2016htt}. Beyond the astrophysics that gravitational waves alone can reveal, a campaign is underway to search for electromagnetic counterparts to binary mergers~\citep{Abbott:2016gcq,Abbott:2016iqz,Abbott:2016nhf}. Joint electromagnetic and gravitational-wave observations will yield a tremendous amount of information about mergers, their host environments, and their ejecta~\citep{Metzger:2011bv}.

The purpose of LIGO's binary merger search is to identify 
astrophysical events in the detectors' noise-dominated data.
Several different methods were used to identify gravitational-wave signals~\citep{Usman:2015kfa,2017PhRvD..95d2001M,Klimenko:2015ypf}
in Advanced LIGO's first observing 
run~\citep{TheLIGOScientific:2016pea,Abbott:2017iws}.
In this paper,
we discuss the PyCBC binary-merger search~\citep{Canton:2014ena,Usman:2015kfa,alex_nitz_2017_545845} which uses matched filtering~\citep{Wiener:1949,Cutler:1992tc,Allen:2005fk} with accurate models of gravitational waveforms~\citep{Blanchet:2002av,Buonanno:1998gg,Faye:2012we,Taracchini:2013rva,Bohe:2016gbl} to 
identify signals. The PyCBC search has been used to the detect binary black hole mergers in the first Advanced LIGO
observing run~\citep{TheLIGOScientific:2016qqj} and to place upper limits
on the rate of binary neutron star and neutron star--black hole mergers~\citep{Abbott:2016ymx}.

Since LIGO detector noise contains both non-stationary and non-Gaussian components, matched filtering will report high signal-to-noise ratios for events that are not astrophysical~\citep{Nuttall:2015dqa,TheLIGOScientific:2016zmo}. To suppress noise, the PyCBC search requires that a source is seen consistently in both LIGO detectors~\citep{Usman:2015kfa} and performs an additional consistency test between the
data and 
the target waveform~\citep{Allen:2004gu}. 
Using the signal-to-noise ratio and the signal-consistency tests, each coincident event is assigned a number determined by the search's \emph{detection statistic}. By construction, the greater the detection statistic's value, the more likely the event is to be an astrophysical signal. 
The detection statistic alone is typically not sufficient to determine if an event is astrophysical, since we do not know the distribution of this statistic due to detector noise; this noise background cannot be accurately modeled and must be empirically measured for the search. To determine an event's significance, we calculate the \emph{false alarm rate} of the search~\citep{TheLIGOScientific:2016qqj}.
The search's false alarm rate measures 
how often it would report a non-astrophysical event with a detection-statistic value as high as a given candidate event. 
The smaller this false alarm rate is, the more likely the candidate event is to be astrophysical.
The PyCBC search consists of two components: a low-latency search that identifies triggers with an estimate of the false alarm rate for rapid follow-up~\citep{Nitz:pycbc_live}, and an offline search that reprocesses the data with additional detector data-quality information and provides the final statement of candidate significance~\citep{Usman:2015kfa}.

There are two ways to improve the sensitivity of the search for binary merger signals: we can increase the measured signal-to-noise ratio for a signal, or we can reduce the false alarm rate of the search at a given signal-to-noise ratio. The signal-to-noise ratio of a merger could be increased by reducing instrumental noise in the detectors~\citep{Martynov:2016fzi} or improving the fit between the model waveforms used in the search and the true gravitational-wave signal. Work to reduce noise towards the Advanced LIGO detectors' design sensitivity (and beyond) is ongoing~\citep{Aasi:2013wya,Evans:2016mbw}. For binary neutron stars, the fit of the waveform models is already very high for detection searches~\citep{Brown:2012qf}, although improvements will help parameter measurement~\citep{Lackey:2014fwa}. For binary black holes and neutron star--black hole binaries, improvements to the theoretical waveforms, and extension of searches to cover  a larger parameter space (e.g. the effects of binary spin-orbit precession and higher-order modes of gravitational-wave emission) are active areas of research. This paper considers the second of the two approaches to improve search sensitivity: improving the separate ability to separate astrophysical signals from detector noise.
Here we restrict our study to the waveforms and search parameter space used in Advanced LIGO's first observing run~\citep{TheLIGOScientific:2016qqj}.

We present a new detection statistic that reduces the false alarm rate of the search across the entire target signal space: binary neutron stars, binary black holes, and neutron star--black hole binaries. 
This detection statistic uses information from both the physical parameters
of the gravitational-wave signal and the expected distribution of the noise 
background to re-scale ``combined effective signal-to-noise ratio'' used in 
previously by the PyCBC search. This is similar to detector-sensitivity 
weighting of the signal-to-noise ratio proposed by 
\cite{Biswas:2012tv,Biswas:2012ty} and used in the Initial LIGO and 
Virgo binary-merger searches \citep{Abadie:2010yb}. 
We test this detection statistic using data from Advanced LIGO's first observing run. 
We explain the meaning of the PyCBC search's false alarm rate and detection statistics, and connect these to the traditional matched filter signal-to-noise ratio and to the source's luminosity distance. Using simulated signals we show that the improvements described here yield a $\sim 10\%$ increase in the detection rate of binary neutron stars and a $\sim 20\%$ increase in the detection rate of neutron star--black hole binaries compared to the search used in the first Advanced LIGO observing run~\citep{TheLIGOScientific:2016qqj,TheLIGOScientific:2016pea}. These techniques are currently in use in binary merger searches in Advanced LIGO's second observing run and was used to evaluate events such as GW170104~\citep{Abbott:2017vtc}. Finally, we propose an additional improvement---not yet implemented---that can increase the binary black hole detection rate by $\sim 30\%$ beyond the search used in Advanced LIGO's first observing run, without negatively impacting the sensitivity for lower-mass mergers.

\section{Measuring Event Significance}
\label{sec:stat}

\begin{figure*}
\centering
\includegraphics[width=0.47\textwidth]{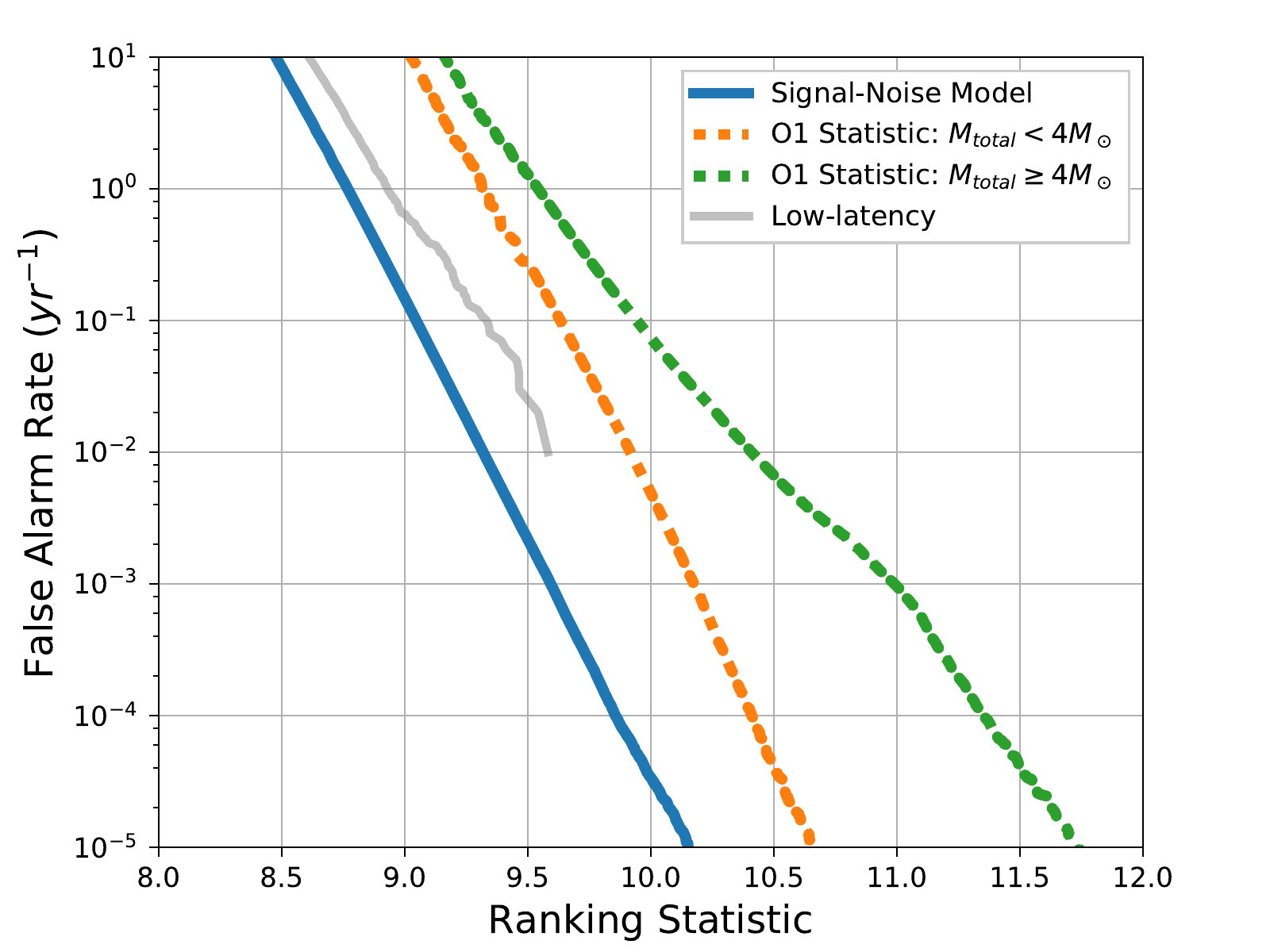}
\includegraphics[width=0.49\textwidth]{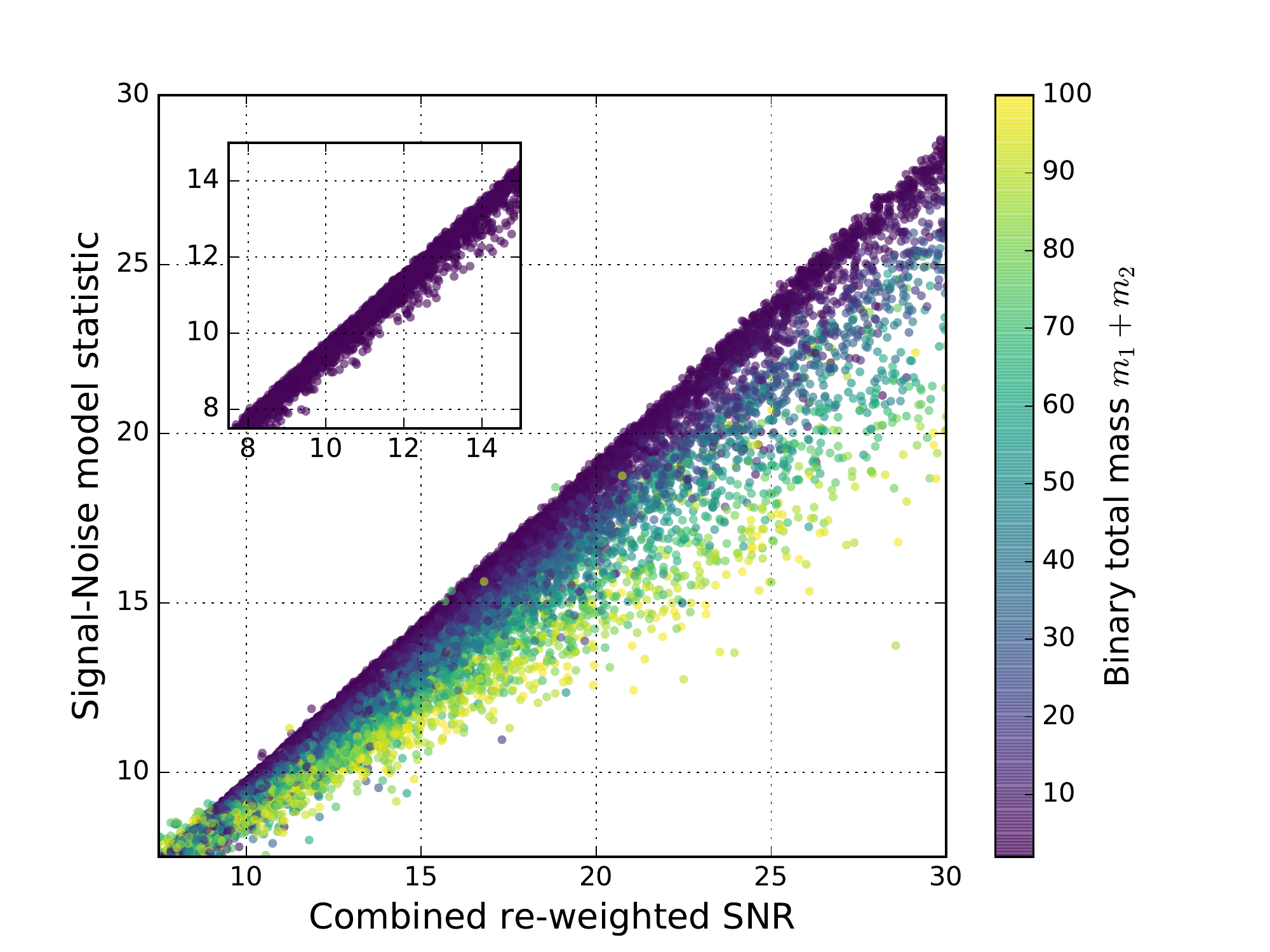}
\caption{%
\label{f:far}
Left: The PyCBC search's false alarm rate measured using data from Advanced LIGO's first observing run. The false alarm rate is shown as a function of the detection statistic used by the search.  The dotted lines show the search's false alarm rate as a function of the combined re-weighted signal-to-noise ratio $\hat{\rho}$ used to detect the binary mergers in Advanced LIGO's first observing run.
The orange line shows this false alarm rate for binary neutron star sources ($M \le 4 M_\odot$) and the green line for binary black holes and neutron star-black hole sources ($M \ge 4 M_\odot$). The solid blue line shows the false alarm rate of the offline PyCBC search using the new signal- and noise-weighted detection statistic $\varrho$ proposed here. The solid grey line shows the false alarm rate of the low-latency PyCBC search, 
designed to rapidly identify triggers for electromagnetic follow-up, at a typical point in time during the first observing run. Note that this curve will vary over time due to changing noise characteristics. The low-latency search currently uses the $\tilde{\rho}$ detection statistic of Eq.~\eqref{eq:online}, which suppresses triggers with suboptimal values of $p^S(\vec{\theta})$, and so will have a reduced detection statistic value at a fixed false alarm rate. 
Right: Using simulated merger signals in data from Advanced LIGO's first observing run, we construct a map between the new detection statistic $\varrho$ and the combined re-weighted signal-to-noise ratio $\hat{\rho}$. The inset shows this map for binary neutron star sources. For these signals, the $\hat{\rho}$, and hence $\varrho$, scale approximately linearly with the inverse luminosity distance to the source. Together, these figures show that our new detection statistic gives around one order of magnitude improvement in the significance of a binary neutron star merger at a given luminosity distance. The improvement for binary black holes depends on the masses of the compact objects with higher mass systems showing less improvement, as indicated by the scatter below the diagonal line seen in the right hand figure.
}
\end{figure*}

If the LIGO detector noise was stationary and Gaussian, a signal time-of-arrival test would suffice to identify candidate events in the detector network, and the quadrature sum of the matched filter signal-to-noise ratio $\rho$ of an event in each Advanced LIGO detector would be an effective detection statistic~\citep{Wainstein:1962,Cutler:1992tc,Pai:2000zt,Allen:2005fk}. In this simple case, we could construct an analytic model of how often the detector background noise produces an event that is as loud as a given event candidate~\citep{Finn:1992wt,Finn:1992xs}; this is the search's false alarm rate. In reality, noise transients and other non-stationarity in the detectors produce excursions in the signal-to-noise ratio that can mimic astrophysical events \emph{and} prevent us from analytically modeling the search's false alarm rate. 

To suppress transient noise, the PyCBC search uses a $\chi^2$ test that checks that the time-frequency evolution of an event matches that of the target waveform~\citep{Allen:2004gu}. This test is constructed so that $\chi^2 \sim 1$ for a real signal and is larger for transient noise. If $\chi^2 \ge 1$ for a trigger, we down-weight the signal-to-noise ratio according to $\hat{\rho} = \rho / [1 + (\chi^2)^3]^{1/6}$~\citep{Babak:2012zx}. The detection statistic used in Advanced LIGO's first observing run was the quadrature sum of the re-weighted signal-to-noise ratio $\hat{\rho}$ observed in each detector, denoted $\hat{\rho}_c$~\citep{TheLIGOScientific:2016qqj}. 

The non-stationarity of the detectors' noise means that it is impossible to construct an analytic model of the search's false alarm rate. We also cannot shield the detectors from gravitational waves and directly measure the signal-free background noise. To ``turn off the signals'' we artificially shift the time\-stamps of one detector's data by an offset that is large compared to the gravitational-wave propagation time between the observatories and produce a new set of coincident events from this time-shifted data set. Real signals will no longer be time-coincident, and the remaining events will be produced by the detector noise alone. This process is repeated with many different time shifts to measure the search's false alarm rate. Under the assumption that signals are relatively rare and that instrumental noise is uncorrelated in time between detectors, this is an effective way to measure the search's false alarm rate.

In practice, the search background varies across the target signal space. Higher mass compact-object mergers produce shorter waveforms which look more like detector glitches and the $\chi^2$ test is less effective at rejecting noise transients in this region of the signal space.  This means that the search's false alarm rate at a given $\hat{\rho}_c$ 
value is not the same for binary neutron stars as it is for binary black holes. In Advanced LIGO's first observing run, candidate and background events were divided into three search classes based on template length. 
The significance of any given event was determined by comparing its detection-statistic value to the false alarm rate in its own class; for a candidate seen in two or more classes, the significance reported is the maximum of the three possible values from the three classes. 
To account for searching over multiple classes, the false alarm rate is increased by a trials factor equal to the number of classes~\citep{Lyons:1900zz}. 

Figure~\ref{f:far} (left) show the false alarm rate of the PyCBC search as a function of the detection statistics considered in this paper, as measured using data from Advanced LIGO's first observing run. 
The dotted lines in Figure~\ref{f:far} show the false alarm rate as a function of $\hat{\rho}_c$, i.e. the detection statistic used in the PyCBC offline search of~\cite{Abbott:2016blz,Abbott:2016nmj,TheLIGOScientific:2016pea,Abbott:2016ymx}. 
The lower (orange) dotted line shows the false alarm rate for binary neutron star sources (defined as mergers with total mass $M = m_1 + m_2 \le 4 M_\odot$), and the upper (green) line shows the false alarm rate for binary mergers with total mass $4 \le M = m_1 + m_2 \le 100$ and a gravitational-wave frequency $f_\mathrm{peak} \ge 100$~Hz at the peak amplitude of the template~\citep{TheLIGOScientific:2016pea}; this class includes both binary black hole and neutron star--black hole sources. 
The search's false alarm rate can be understood as follows: For a given astrophysical source, 
the signal-to-noise ratio scales linearly as the inverse of the distance $\rho \propto 1/d_L$. Since, by construction, $\hat{\rho} \sim \rho$ for real signals, it follows that $\hat{\rho} \propto 1/d_L$ for real signals.  As $d_L$ decreases, the detection statistic increases and the false alarm rate is reduced according to the curves shown in Figure~\ref{f:far}; the event is therefore more significant.

The solid lines of Figure~\ref{f:far} (left) show the search false alarm rate for the new detection statistics proposed here measured using the same data from Advanced LIGO's first observing run. The dark (blue) line shows the full statistic used by the PyCBC offline search and the light (grey) line shows the false alarm rate of the PyCBC low-latency search. These results differ as the offline search uses a large amount of data to measure the noise background, whereas the low-latency search only uses the previous five hours of data. Consequently, the offline search can measure false alarm rates to a precision of better than 1 in 10,000 years, whereas the low-latency search measures the false alarm rate to 1 in 100 years. This is sufficient to identify events for electromagnetic follow-up, with the final event significance measured by the offline search. The low-latency search also uses a simplification of the full detection statistic proposed here, as described in Section~\ref{s:stat} below.

A key result of Figure~\ref{f:far} is that \emph{for all of these searches} the the false alarm rate is an extremely steep function of the detection statistic value. To illustrate the effect of the steep change in false alarm rate, consider a hypothetical binary neutron star source in Advanced LIGO's first observing run. The signal-to-noise ratio of the source depends on the orientation and sky location of the source. Suppose a source was oriented so that it produced a signal-to-noise ratio of $7$ in each LIGO detector. In the first observing run, the average luminosity distance to such a source would be $d_L \sim 85$~Mpc, but it could be as far away as $d_L \sim 194$~Mpc, if it were favorably oriented. The combined detection statistic for this source would be $\hat{\rho}_c = 9.9$. 
For a binary neutron star signal this loud, the search's false alarm rate is $\sim 10^{-2}$~yr$^{-1}$, meaning that noise is expected to produce an event with $\hat{\rho}_c \geq 9.9$ in the $M \le 4 M_\odot$ class once every $\sim 300$ years, which becomes once every 100 years after accounting for the trials factor.
If the source's luminosity distance was increased by $10\%$, then $\hat{\rho}_c$ decreases to $9$: at this statistic value the search's false alarm rate is $\sim 10$~yr$^{-1}$. Since the noise background drops very quickly as a function of signal amplitude, the edge of the observable signal space is very sharp in signal amplitude. 

Figure~\ref{f:far} also shows that small changes in the detection-statistic value due to noise can have a large effect on the measured false alarm rate of a signal. This can cause two sources with similar strain amplitudes to have significant differences between their false alarm rates due to fluctuations in the detector noise at the time of each event. A change of half a unit in signal-to-noise ratio can result in a order of magnitude difference in the measured false alarm rate of a signal.
Similarly, slight differences in implementation between low-latency and offline analyses (e.g.  in the computation of the noise power spectral density, or in the number of $\chi^2$-veto bins) can cause substantial differences in false alarm rate. Highly significant signals, for example GW150914 and GW151226, will be robust to such changes, but the false alarm rate of marginal signals may change significantly.

\section{Improving Search Sensitivity}
\label{s:stat}

Previously, the PyCBC search used four quantities to construct its detection statistic: the matched filter signal-to-noise ratio and the value of the $\chi^2$ test in each of two detectors. However, the search also records the difference in time of arrival between the LIGO Hanford and Livingston observatories $\delta t = t_H - t_L$, and the difference in the phase of the gravitational waveform $\delta \phi = \phi_H - \phi_L$. These time and phase differences, as well as the signal-to-noise ratio in each detector $\rho_{H,L}$, depend on location of the source relative to the detectors. Therefore an astrophysical population of signals will have a nontrivial distribution of events over the parameters $(\rho_H, \rho_L, \delta\phi, \delta t)$. However, under the assumption that noise is not correlated between the LIGO Hanford and Livingston detectors, events due to background noise will be distributed uniformly in $(\delta\phi, \delta t)$; this information can be used to improve the detection statistic.

We can also use the measured noise background as a function of the template parameters to account for the variation of the search background over the target signal space. In 
the search in LIGO's first observing run, the search space was divided the signal space into three classes: 
this provides a crude accounting of the mass-dependent variation in the search background. Here, we show that it is possible to construct a better model of the noise background over the search space, as a function of the masses and spins $(m_1, m_2, \vec{s}_1, \vec{s}_2)$ of the merging compact objects.

We describe a new detection statistic that uses the full information $\vec{\theta} = (\rho_H, \rho_L, \chi^2_H, \chi^2_L, \delta\phi, \delta t, m_1, m_2, \vec{s}_1, \vec{s}_2)$ for a candidate event that significantly improves the sensitivity of the PyCBC search. We construct this statistic by approximating the probability densities of both signal events and noise events over these parameters and forming the \emph{ratio of densities}.  This is the equivalent of the full likelihood ratio for a reduced data set consisting of only the signal-to-noise ratio maxima. A general detection statistic resulting from approximations to the density of signal $p^S(\vec{\theta})$ and noise $p^N(\vec{\theta})$ over the parameter space $\vec{\theta}$ of coincident events can be written as\footnote{We may rescale the detection statistic by a constant (positive) factor and add a constant without affecting the relative ranking of events.}
\begin{equation}\label{eq:genstat}
  \varrho^2 \propto 2 \left[ \log p^S(\vec{\theta}) - \log p^N(\vec{\theta}) 
  \right] + \mathrm{constant.}
\end{equation}
We choose this form for the detection statistic since in the case that the detectors' noise was stationary and Gaussian we have $p^N(\vec{\theta}) \sim\exp[-(\rho_H^2+ \rho_L^2)/2]$ and we recover the standard quadrature sum signal-to-noise ratio statistic. 
\begin{figure}
\centering
\includegraphics[width=0.49\textwidth]{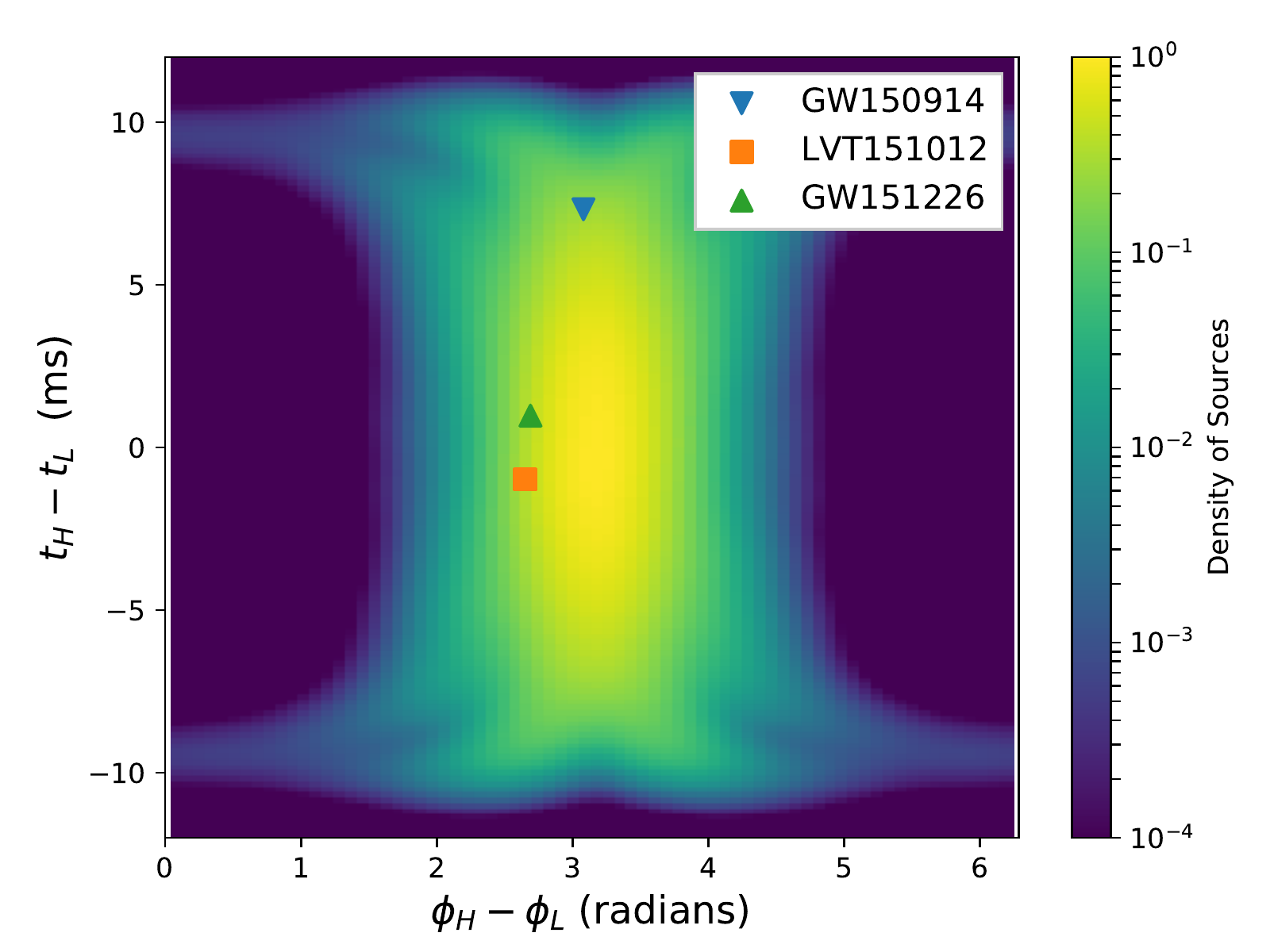}
\caption{The unnormalized distribution of phase differences and time delays between the LIGO Hanford and Livingston Observatories for a simulated population of sources isotropically distributed over the sky and uniformly distributed in space. Noise events have a uniform distribution of time delays and phase differences.  The events from LIGO's first observing run are overlaid; they are consistent with the signal distribution.
\label{f:phasetime}}
\end{figure}

We first consider the dependence on $\delta\phi$ and $\delta t$. For a single detector, the expected distribution of signals over $\phi, t$ is
uniform, and thus does not aid in separating signals from noise.  In a multi-detector network the phase and time \emph{differences} between detectors are determined by a source's sky location and orientation in relation to the detector locations. However, the distribution of noise events will be uniform in $\delta t$ and $\delta \phi$. To use this information, we must first construct the probability distribution in this parameter space for coincident signals from an  astrophysical population. We perform a Monte-Carlo simulation with a population of binary mergers that is uniform in spatial volume and isotropic over the sky and over source orientation. We use the antenna patterns of the detectors located at the LIGO Hanford and Livingston Observatories to calculate the expected signal-to-noise ratio of a source in each detector. If the single detector signal-to-noise ratio falls below a threshold imposed in the search (a value of $\rho = 5.5$ has been used in recent searches), the source is considered not detectable. For each of the remaining sources we record $\delta t$, $\delta \phi$, $\rho_H$, and $\rho_L$, building up a multi-dimensional distribution.  We account for a possible difference in sensitivity between the detectors by repeating the simulation over a range of relative detector sensitivities. The resulting multi-dimensional histogram is then smoothed with a Gaussian kernel whose width is determined by the expected measurement errors in each parameter, to obtain our approximation of the signal distribution $p^S(\vec{\theta})$. This Monte-Carlo simulation is performed once for each detector network and stored for future use as an efficient look-up table.

In Figure~\ref{f:phasetime} we show the resulting (unnormalized) signal distribution over the two most important parameters, the Livingston-Hanford time delays and phase differences, having marginalized over the parameters not shown.  While the expected background distribution is uniform over $\delta t$ and $\delta \phi$, the signal distribution clearly is not. This information about $p^S(\vec{\theta})$ can improve the separation of signal from noise events and thus increase search sensitivity. The peaking of the phase difference around $\pi$ is due to the relative positions and orientations of the two LIGO detectors, which are close to coplanar and have a $\sim 90^\circ$ relative rotation angle of the interferometer arms.  For most sources they observe approximately the same polarization with opposite sign of the gravitational-wave waveform. However, for arrival directions close to the line joining the two detectors, depending on the polarization, there may be partial cancellation or reversal of sign in the detector responses allowing a full range of relative phases between $0$--$2\pi$.
The phase difference distribution will in general be different for other combinations of observatories. We can construct an improved detection statistic using this information, defined as
\begin{equation}\label{eq:online}
 \tilde{\rho}^2 = \hat{\rho}_c^2 + 2 \log\left(\frac{p^S(\vec{\theta})}
 {p^S_{\max}}\right),
\end{equation}

where $p^S_{\max}$ is the most likely (highest) value in the multi-dimensional histogram. This is the detection statistic used by the PyCBC low-latency search~\citep{Nitz:pycbc_live}. 
\begin{figure*}
  \centering
      \includegraphics[width=\textwidth]{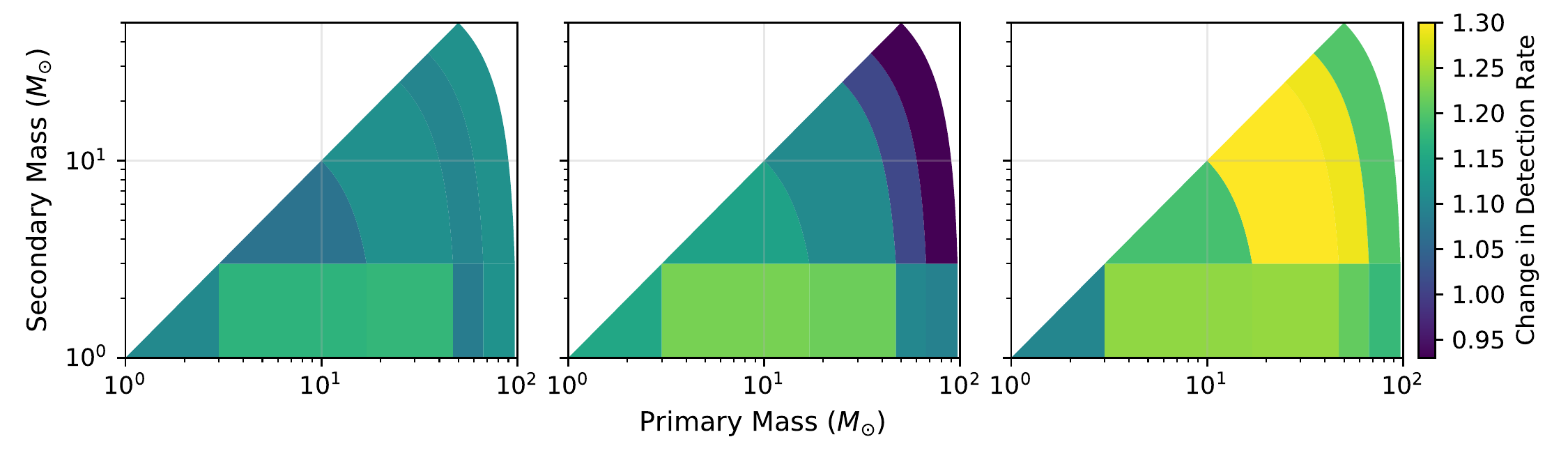}
  \caption{\label{vt}
Comparison of the expected rate of detections with the PyCBC pipeline using 
  different detection statistics, estimated via a set of simulated signals added to 
  Advanced LIGO data.  Comparisons are relative to the statistic used in the first 
  observing run.  
  Left: with a statistic using information on source sky location and orientation   
   sensitivity increases for all sources.
  Center: including also an improved model of the background distribution, the excess noise in higher-mass templates is taken into 
  account and we see further sensitivity increases for BNS and NSBH sources.
  Right: using a mass prior uniform in the logarithm of component masses 
   sensitivity is significantly increased for higher-mass
  sources.
  \vspace*{0.2cm}
  }
\end{figure*}

So far we are still considering an idealized noise distribution, $p^N \sim \exp[-(\hat{\rho}_{L}^2 + \hat{\rho}_{H}^2)/2]$, which neglects how the background may vary with the target waveform parameters and/or between detectors. 
However, the noise distributions in real data are quite different across different templates and between different detectors~\citep{TheLIGOScientific:2016qqj}. The previous method of dividing the search into 
separate background classes has the drawbacks of requiring a somewhat arbitrary choice of class boundaries; failing to model background variation within a class, thus potentially hurting sensitivity to some types of signal; and, if a large number of classes are used, a high trials factor diluting the significance of an event in the search as a whole compared to an estimate from a single class.

We obtain a more accurate and flexible description over the whole search space via an explicit model of the rate of occurrence of single-detector signal-to-noise ratio maxima due to noise:
\begin{equation}
\label{eq:single_noise}
 \lambda^N_{Aw}(\hat{\rho}_A) = \mu_{Aw} \alpha_{Aw} \exp[-\alpha_{Aw}
 (\hat{\rho}_A - \hat{\rho}_{\rm th})],
\end{equation}
where $\mu^N_{Aw}$ is the total noise event rate in a given template $w$ and detector $A$ having $\hat{\rho}$ above a threshold value $\hat{\rho}_{\rm th}$, and $\alpha_{Aw}$ describes how steeply the noise distribution falls off.  The values of $\mu$ and $\alpha$ in each template and detector are empirically determined by a two-step procedure: first, a maximum-likelihood fit is performed to the signal-to-noise ratio maxima in each template and detector, which are assumed to be dominated by noise. The distribution at large $\hat{\rho}$ values is likely to be dominated by signals; hence, when performing the fit we remove a predetermined number of the highest-$\hat{\rho}$ maxima.  Then to reduce variance due to small number statistics, we perform kernel smoothing of the $\lambda$ and $\mu$ values, in each detector, over the duration of the template in the LIGO sensitive band $\tau_w$, since templates with similar $\tau$ have similar responses to transient noise artifacts. 

We expect noise events to be uncorrelated in time between different detectors: thus, the expected distribution of coincident noise over $\hat{\rho}_{H,L}$ in a given template is the product of single-detector distributions.  For a template of duration $\tau$ we have \begin{equation}
\label{eq:coinc_noise}
 p^N(\vec{\theta}) \propto \lambda^N_H(\hat{\rho}_H,\tau)
 \lambda^N_L(\hat{\rho}_L,\tau), 
\end{equation}
where each factor on the right hand side is given by the exponential model of Eq.~\eqref{eq:single_noise} using the smoothed fit parameters $\mu_A(\tau)$, $\alpha_A(\tau)$ ($A=H,L$).

Using this approximation for the noise distribution in the general statistic Eq.~\eqref{eq:genstat} has the effect of smoothly rescaling the statistic values of events depending on the steepness of the noise distributions in the template where the events appear.  Figure~\ref{f:far} (right) shows the new statistic value for a set of simulated signal events added to O1 data.  We use the freedom to rescale the statistic value by a constant factor to set the values for low-mass candidate events, recovered in templates with steep background distributions, to be similar to $\hat{\rho}_c$.  Then given the elevated rate of high-$\hat{\rho}$ noise events in higher-mass templates, the statistic values are suppressed for more massive candidates. In Figure~\ref{f:far} (left), we see that the background classes used in the first observing run are combined into a single noise distribution, with a noise level significantly suppressed relative to the search used in O1, even for binary neutron star templates where the $\chi^2$ test is most effective at suppressing noise artefacts. This statistic is used in the PyCBC offline search in Advanced LIGO's second observing run.

We have not yet included the parameters of the waveform template (binary component masses and spins) in the signal part of our detection statistic. Not considering these parameters is only an optimal choice if the distribution of signal events over them, at a given $\rho$, is approximately equal to the distribution of waveform templates~\citep{Dent:2013cva}. In binary merger searches, we use a discrete set of templates laid out to bound the maximum possible loss of detection rate for any of our target sources~\citep{Capano:2016dsf,Brown:2012qf,Ajith:2012mn}.  The density of noise events over the parameter space of masses and spins varies as the density of templates: due to the way the binary parameters affect the waveform, this density varies by many orders of magnitude between low- and high-mass regions.  The statistics described above then have an implicit prior that the rate of detectable signals is much higher in the low-mass region -- although they certainly allow the detection of high-mass signals. 

We can introduce an explicit model of the mass (and/or spin) distribution by including a term in the general statistic Eq.~\eqref{eq:genstat} which depends on the binary parameters of the template waveform $w$.  This term is the logarithmic ratio of the expected density of signals over the parameter space to the density of templates over the same space. Here we neglect statistical errors in the template parameters compared to the true  parameters of the merging system; this approximation is likely to be valid for signal distributions that do not vary very abruptly over the binary system's parameters~\citep{Dent:2013cva}. As an example of a possible prior, we take a distribution of sources which is uniform over the logarithms of binary component masses; the ratio of template density to signal density is found via a simple kernel density estimate applied to the set of template $(\log m_1, \log m_2)$ values.

The relative rates of signals arriving at Earth from binary mergers of different masses are still highly uncertain~\citep{Abadie:2010cf}, so applying a complex or fully realistic model prior is not appropriate.  Note that a prior might be chosen to maximize either the total expected number of detections, or some other figure of merit for the results of a search. 

\section{Improvement in Detection Rate and Significance}

In this section we evaluate how the use of the improved detection statistics described 
in the previous sections affects the sensitivity of searches for binary neutron star, 
neutron-star black-hole and binary black hole mergers. We simulate a population of 
sources and insert these into data from Advanced LIGO's first observing run. Using the 
PyCBC analysis to recover these sources, we determine which would be significant
candidates, corresponding to a false alarm rate of less than 1 per 100 years.
The volume of space that the search is sensitive to, which is proportional to the 
expected number of detections, is then obtained by Monte Carlo integration~\citep{Usman:2015kfa}.

We simulate an isotropic distribution of sources in sky location, orientation, and spin 
angle, i.e.\ orientation of the binary component angular momenta relative to the orbital
angular momentum; thus, the binaries have generically precessing orbital dynamics. The 
magnitude of the spin vector is restricted to $<0.4$ for source with component masses 
less than $3\,M_\odot$, and $<0.989$ otherwise. We consider a population of sources
that is uniformly distributed in component mass.
Compared to the 
detection statistic used in the first observing run~\citep{TheLIGOScientific:2016pea}, we show 
in Fig~\ref{vt} how sensitivity is improved using our new statistics 
that take into account 
(left) the location and orientation of sources, (center) a more accurate background
model, and (right) a prior over the source mass distribution uniform in $(\log m_1,
\log m_2)$.  We achieve a $\sim10\%$ and $\sim20\%$ increase in the number of detections 
of binary neutron star mergers and neutron star--black hole mergers, respectively. 

An explicit prior on the mass distribution of detectable sources is not currently being
employed in the analysis of Advanced LIGO data; however if introduced, a $\sim 30\%$ 
improvement in the rate of detections for binary black hole mergers with masses similar to those 
identified in \citep{TheLIGOScientific:2016pea}
could be achieved while maintaining or improving upon the sensitivity to binary neutron star and neutron star--black hole mergers.

\section{Conclusions}

Although we have described relatively straightforward improvements to our search detection 
statistic, we may still not have achieved an optimal event ranking: various issues remain
to be addressed by future work.  We have approximated the signal and noise event rates as
constant (stationary) over time for the duration of each experiment; however, despite the 
application of data quality vetoes \citep{Nuttall:2015dqa,TheLIGOScientific:2016zmo} to 
remove times of known environmental and instrumental disturbances from the search, we
occasionally observe large upward fluctuations in noise event rate on the scale of seconds
to minutes due to non-stationary or non-Gaussian data.  Search efficiency could be further
improved by down-ranking these times via a model incorporating a time-dependent noise 
rate.  Also, it may be possible to develop other diagnostics besides the $\chi^2$ 
test~\citep{Allen:2004gu} to distinguish signals from noise transients
in single-detector data. 

The reader may ask what the effect these detection statistics have on the significance of the candidate events seen in O1. GW150914 and GW151226 have high enough SNR that
their statistic values would lie above all available background samples for any of the
statistics we present, so we can only place a bound on their significance in all cases. For the marginal event LVT151012 event, its phase difference, time difference and amplitude parameters lie well within the expected signal distribution. 
An evaluation of its significance using the signal-noise model ranking statistic $\varrho$ would require a re-analysis of the O1 data which is beyond the scope of this paper.  To estimate the possible reduction in false alarm rate we may use the expected signal distribution of Section~\ref{sec:stat} 
to find the relative likelihood of a signal event with a specific $\delta\phi$ and $\delta t$ value as compared to a noise event.  As in Fig.~\ref{f:phasetime} we marginalize over the other dimensions of the signal distribution, and note that noise events are uniformly distributed over $\delta\phi,\,\delta t$.  At the values measured for LVT151012, we find that the density (normalized pdf) of signal events is $\sim 10$ times higher than that of noise events, suggesting that if this information were included in the ranking statistic of the search the event would be assigned a false alarm rate lower by approximately an order of magnitude. 

The significance of marginal events is also very sensitive to the relative weighting assigned to different regions of parameter space.  As an example, LVT151012 would have been assigned a false alarm rate of less than 1 per 1000 years if the proposed logarithmic-in-component-mass prior was incorporated into the ranking statistic.  We warn, though, that a nontrivial mass prior such as proposed at the end of Section~ref{s:stat} is not an unbiased choice with respect to previously identified events such as LVT151012.

In this study we have considered template waveforms for dominant-mode gravitational-wave emission from 
non-precessing (and quasi-circular) binaries.  Such templates are in general not expected 
to have a uniformly optimal detection efficiency for more complex binary merger signals: in 
particular those including non-dominant modes~\citep{Capano:2013raa}, relevant for high-mass mergers 
with significantly unequal component masses, and those showing strong
amplitude modulation as a result of spin-orbital precession~\citep{Harry:2016ijz,Harry:2013tca}.  To best detect such
signals, templates with more degrees of freedom may be required~\citep{Capano:2013raa,Harry:2013tca}, implying that each 
candidate event may be described by numerous independent matched filter outputs.  
Optimizing the efficiency of searches over a larger event parameter space will likely
require modeling the expected distributions of signal and noise over that space: 
techniques similar to those presented here will be applicable to this task. 

\acknowledgments
The authors thank the LIGO Scientific Collaboration for access to the data and gratefully acknowledge the support of the United States National Science Foundation (NSF) for the construction and operation of the
LIGO Laboratory and Advanced LIGO as well as the Science and Technology 
Facilities Council (STFC) of the United Kingdom, and the Max-Planck-Society 
(MPS) for support of the construction of Advanced LIGO. Additional support 
for Advanced LIGO was provided by the Australian Research Council.
We thank Christopher Biwer, Soumi De, Ian Harry, Greg Mendell, Frank Ohme, Peter Saulson, Bernard Schutz, and Alan Weinstein for helpful discussions. TDC is supported by an appointment to the NASA Postdoctoral Program at the Goddard Space Flight Center, administered by Universities Space Research Association under contract with NASA. DAB acknowledges support from National Science Foundation awards PHY-1404395, ACI-1443047. DAB performed part of this work at the Kavli Institute for Theoretical Physics, which is supported by NSF award PHY-1125915.  SF acknowledges support from STFC awards ST/L000962/1. We thank the Atlas cluster computing team at AEI Hannover where this analysis was carried out. 

\bibliographystyle{apj}
\bibliography{references}

\begin{thebibliography}{}
\expandafter\ifx\csname natexlab\endcsname\relax\def\natexlab#1{#1}\fi

\bibitem[{Aasi {et~al.}(2016)}]{Aasi:2013wya}
Aasi, J., {et~al.} 2016, Living Rev. Rel., 19, 1

\bibitem[{Abadie {et~al.}(2010{\natexlab{a}})}]{Abadie:2010cf}
Abadie, J., {et~al.} 2010{\natexlab{a}}, Class. Quant. Grav., 27, 173001

\bibitem[{Abadie {et~al.}(2010{\natexlab{b}})}]{Abadie:2010yb}
---. 2010{\natexlab{b}}, Phys. Rev., D82, 102001

\bibitem[{Abbott {et~al.}(2016{\natexlab{a}})}]{TheLIGOScientific:2016htt}
Abbott, B.~P., {et~al.} 2016{\natexlab{a}}, Astrophys. J., 818, L22

\bibitem[{Abbott {et~al.}(2016{\natexlab{b}})}]{TheLIGOScientific:2016pea}
---. 2016{\natexlab{b}}, Phys. Rev., X6, 041015

\bibitem[{Abbott {et~al.}(2016{\natexlab{c}})}]{TheLIGOScientific:2016zmo}
---. 2016{\natexlab{c}}, Class. Quant. Grav., 33, 134001

\bibitem[{Abbott {et~al.}(2016{\natexlab{d}})}]{TheLIGOScientific:2016qqj}
---. 2016{\natexlab{d}}, Phys. Rev., D93, 122003

\bibitem[{Abbott {et~al.}(2016{\natexlab{e}})}]{TheLIGOScientific:2016agk}
---. 2016{\natexlab{e}}, Phys. Rev. Lett., 116, 131103

\bibitem[{Abbott {et~al.}(2016{\natexlab{f}})}]{Abbott:2016nmj}
---. 2016{\natexlab{f}}, Phys. Rev. Lett., 116, 241103

\bibitem[{Abbott {et~al.}(2016{\natexlab{g}})}]{Abbott:2016gcq}
---. 2016{\natexlab{g}}, Astrophys. J., 826, L13

\bibitem[{Abbott {et~al.}(2016{\natexlab{h}})}]{Abbott:2016blz}
---. 2016{\natexlab{h}}, Phys. Rev. Lett., 116, 061102

\bibitem[{Abbott {et~al.}(2016{\natexlab{i}})}]{Abbott:2016iqz}
---. 2016{\natexlab{i}}, Astrophys. J. Suppl., 225, 8

\bibitem[{Abbott {et~al.}(2016{\natexlab{j}})}]{Abbott:2016nhf}
---. 2016{\natexlab{j}}, Astrophys. J., 833, L1

\bibitem[{Abbott {et~al.}(2016{\natexlab{k}})}]{Abbott:2016ymx}
---. 2016{\natexlab{k}}, Astrophys. J., 832, L21

\bibitem[{Abbott {et~al.}(2017{\natexlab{a}})}]{Evans:2016mbw}
---. 2017{\natexlab{a}}, Class. Quant. Grav., 34, 044001

\bibitem[{Abbott {et~al.}(2017{\natexlab{b}})}]{Abbott:2017vtc}
---. 2017{\natexlab{b}}, Phys. Rev. Lett., 118, 221101

\bibitem[{Abbott {et~al.}(2017{\natexlab{c}})}]{Abbott:2017iws}
---. 2017{\natexlab{c}}, Phys. Rev., D96, 022001

\bibitem[{Ajith {et~al.}(2014)Ajith, Fotopoulos, Privitera, Neunzert, \&
  Weinstein}]{Ajith:2012mn}
Ajith, P., Fotopoulos, N., Privitera, S., Neunzert, A., \& Weinstein, A.~J.
  2014, Phys. Rev., D89, 084041

\bibitem[{Allen(2005)}]{Allen:2004gu}
Allen, B. 2005, Phys. Rev. D, 71, 062001

\bibitem[{Allen {et~al.}(2012)Allen, Anderson, Brady, Brown, \&
  Creighton}]{Allen:2005fk}
Allen, B., Anderson, W.~G., Brady, P.~R., Brown, D.~A., \& Creighton, J. D.~E.
  2012, Phys.Rev., D85, 122006

\bibitem[{Babak {et~al.}(2013)Babak, Biswas, Brady, Brown, Cannon,
  {et~al.}}]{Babak:2012zx}
Babak, S., Biswas, R., Brady, P., {et~al.} 2013, Phys.Rev., D87, 024033

\bibitem[{Biswas {et~al.}(2012{\natexlab{a}})}]{Biswas:2012ty}
Biswas, R., {et~al.} 2012{\natexlab{a}}, Phys. Rev., D85, 122009

\bibitem[{Biswas {et~al.}(2012{\natexlab{b}})}]{Biswas:2012tv}
---. 2012{\natexlab{b}}, Phys. Rev., D85, 122008

\bibitem[{Blanchet(2002)}]{Blanchet:2002av}
Blanchet, L. 2002, Living Rev. Rel., 5, 3

\bibitem[{{Boh{\' e}} {et~al.}(2016)}]{Bohe:2016gbl}
{Boh{\' e}}, A., {et~al.} 2016, arXiv:1611.03703

\bibitem[{Brown {et~al.}(2012)Brown, Harry, Lundgren, \& Nitz}]{Brown:2012qf}
Brown, D.~A., Harry, I., Lundgren, A., \& Nitz, A.~H. 2012, Phys. Rev., D86,
  084017

\bibitem[{Buonanno \& Damour(1999)}]{Buonanno:1998gg}
Buonanno, A., \& Damour, T. 1999, Phys. Rev., D59, 084006

\bibitem[{Capano {et~al.}(2016)Capano, Harry, Privitera, \&
  Buonanno}]{Capano:2016dsf}
Capano, C., Harry, I., Privitera, S., \& Buonanno, A. 2016, Phys. Rev., D93,
  124007

\bibitem[{Capano {et~al.}(2014)Capano, Pan, \& Buonanno}]{Capano:2013raa}
Capano, C., Pan, Y., \& Buonanno, A. 2014, Phys. Rev., D89, 102003

\bibitem[{Cutler {et~al.}(1993)Cutler, Apostolatos, Bildsten, Finn, Flanagan,
  Kennefick, Markovic, Ori, Poisson, Sussman, \& Thorne}]{Cutler:1992tc}
Cutler, C., Apostolatos, T.~A., Bildsten, L., {et~al.} 1993, Phys.Rev.Lett.,
  70, 2984

\bibitem[{Dal~Canton {et~al.}(2014)Dal~Canton, Nitz, Lundgren, Nielsen, Brown,
  Dent, Harry, Krishnan, Miller, Wette, Wiesner, \& Willis}]{Canton:2014ena}
Dal~Canton, T., Nitz, A.~H., Lundgren, A.~P., {et~al.} 2014, Phys. Rev., D90,
  082004

\bibitem[{Dent \& Veitch(2014)}]{Dent:2013cva}
Dent, T., \& Veitch, J. 2014, Phys. Rev., D89, 062002

\bibitem[{Faye {et~al.}(2012)Faye, Marsat, Blanchet, \& Iyer}]{Faye:2012we}
Faye, G., Marsat, S., Blanchet, L., \& Iyer, B.~R. 2012, Class. Quant. Grav.,
  29, 175004

\bibitem[{Finn(1992)}]{Finn:1992wt}
Finn, L.~S. 1992, Phys. Rev., D46, 5236

\bibitem[{Finn \& Chernoff(1993)}]{Finn:1992xs}
Finn, L.~S., \& Chernoff, D.~F. 1993, Phys. Rev., D47, 2198

\bibitem[{Harry {et~al.}(2016)Harry, Privitera, Boh\'e, \&
  Buonanno}]{Harry:2016ijz}
Harry, I., Privitera, S., Boh\'e, A., \& Buonanno, A. 2016, Phys. Rev., D94,
  024012

\bibitem[{Harry {et~al.}(2014)Harry, Nitz, Brown, Lundgren, Ochsner, \&
  Keppel}]{Harry:2013tca}
Harry, I.~W., Nitz, A.~H., Brown, D.~A., {et~al.} 2014, Phys. Rev., D89, 024010

\bibitem[{Klimenko {et~al.}(2016)}]{Klimenko:2015ypf}
Klimenko, S., {et~al.} 2016, Phys. Rev., D93, 042004

\bibitem[{Lackey \& Wade(2015)}]{Lackey:2014fwa}
Lackey, B.~D., \& Wade, L. 2015, Phys. Rev., D91, 043002

\bibitem[{Lyons(2008)}]{Lyons:1900zz}
Lyons, L. 2008, Ann. Appl. Stat., 2, 887

\bibitem[{Martynov {et~al.}(2016)}]{Martynov:2016fzi}
Martynov, D.~V., {et~al.} 2016, Phys. Rev., D93, 112004

\bibitem[{{Messick} {et~al.}(2017){Messick}, {Blackburn}, {Brady}, {Brockill},
  {Cannon}, {Cariou}, {Caudill}, {Chamberlin}, {Creighton}, {Everett}, {Hanna},
  {Keppel}, {Lang}, {Li}, {Meacher}, {Nielsen}, {Pankow}, {Privitera}, {Qi},
  {Sachdev}, {Sadeghian}, {Singer}, {Thomas}, {Wade}, {Wade}, {Weinstein}, \&
  {Wiesner}}]{2017PhRvD..95d2001M}
{Messick}, C., {Blackburn}, K., {Brady}, P., {et~al.} 2017, Phys. Rev., D95,
  042001

\bibitem[{Metzger \& Berger(2012)}]{Metzger:2011bv}
Metzger, B.~D., \& Berger, E. 2012, Astrophys. J., 746, 48

\bibitem[{Nitz {et~al.}(2017{\natexlab{a}})}]{Nitz:pycbc_live}
Nitz, A.~H., {et~al.} 2017{\natexlab{a}}, in preparation

\bibitem[{Nitz {et~al.}(2017{\natexlab{b}})Nitz, Harry, Biwer, Brown, Willis,
  Canton, Pekowsky, Dent, Williamson, Capano, De, Kumar, Machenschalk, Cabero,
  Massinger, Lenon, Fairhurst, Reyes, Nielsen, Kapadia, Pannarale, Singer,
  Babak, Macleod, Sugar, na~Zertuche, Veitch, Couvares, Bockelman, \&
  Brown}]{alex_nitz_2017_545845}
Nitz, A.~H., Harry, I., Biwer, C.~M., {et~al.} 2017{\natexlab{b}}, PyCBC
  Software, doi:10.5281/zenodo.545845

\bibitem[{Nuttall {et~al.}(2015)Nuttall, Massinger, Areeda, Betzwieser, Dwyer,
  Effler, Fisher, Fritschel, Kissel, Lundgren, Macleod, Martynov, McIver,
  Mullavey, Sigg, Smith, Vajente, Williamson, \& Wipf}]{Nuttall:2015dqa}
Nuttall, L.~K., Massinger, T.~J., Areeda, J., {et~al.} 2015, Class. Quant.
  Grav., 32, 245005

\bibitem[{Pai {et~al.}(2001)Pai, Dhurandhar, \& Bose}]{Pai:2000zt}
Pai, A., Dhurandhar, S., \& Bose, S. 2001, Phys. Rev., D64, 042004

\bibitem[{Taracchini {et~al.}(2014)Taracchini, Buonanno, Pan, Hinderer, Boyle,
  Hemberger, Kidder, Lovelace, Mroue, Pfeiffer, Scheel, Szilagyi, Taylor, \&
  Zenginoglu}]{Taracchini:2013rva}
Taracchini, A., Buonanno, A., Pan, Y., {et~al.} 2014, Phys. Rev., D89, 061502

\bibitem[{Usman {et~al.}(2016)Usman, Nitz, Harry, Biwer, Brown,
  {et~al.}}]{Usman:2015kfa}
Usman, S.~A., Nitz, A.~H., Harry, I.~W., {et~al.} 2016, Class. Quant. Grav.,
  33, 215004

\bibitem[{Wainstein \& Zubakov(1962)}]{Wainstein:1962}
Wainstein, L.~A., \& Zubakov, V.~D. 1962, Extraction of signals from noise (New
  Jersey: Prentice-Hall)

\bibitem[{Wiener(1949)}]{Wiener:1949}
Wiener, N. 1949, Extrapolation, Interpolation, and Smoothing of Stationary Time
  Series (New York: Wiley)

\end{thebibliography}

\end{document}